\documentclass[11pt]{article}

\usepackage[final]{acl}
\usepackage{amsmath}
\usepackage{times}
\usepackage{latexsym}

\usepackage[T1]{fontenc}
\usepackage[utf8]{inputenc}

\usepackage{microtype}

\usepackage{inconsolata}

\usepackage{graphicx}

\title{$\pi$-RAG: Oblivious Retrieval via Semantic Quantization and Transcendental Addressing for Large Language Models}

\author{Aniket Wattamwar \\
  \texttt{aniket.wattamwar17@gmail.com} \\\And
  Mrunal Kakirwar \\
  \texttt{kakirwarm@gmail.com} \\}

\begin{document}
\maketitle
\begin{abstract}
% $\pi$
This paper introduces $\pi$-RAG, a novel  architecture for oblivious retrieval that decouples Large Language Models (LLMs) from sensitive data storage without sacrificing semantic understanding. Traditional Retrieval-Augmented Generation (RAG) architectures expose raw vector embeddings to potential inversion attacks and nondeterministic retrieval failures. To address this, we utilize the digits of $\pi$ as a source of transcendental entropy, creating an immutable indirection layer between the LLM and private records. The value $\pi$ provides immutability, is uneditable and math governs it.  The architecture also introduces a Semantic Quantization Layer. This layer projects user inputs onto a pre-computed manifold of Canonical Intent Centroids. RAG performs vector cosine similarity but here it maps the centroids to deterministic offsets via cryptographic salt. The resulting $\pi$-key is a pointer to standardized payload from the actual datastore. By replacing direct access to the datastore via LLM with this transcendental layer, $\pi$-RAG mathematically guarantees that the inference remains oblivious to the data. This architecture unifies deterministic randomness, auditability, and differential privacy, demonstrating high efficacy for high-compliance sectors such as finance and healthcare.
\end{abstract}

\section{Introduction}

With the advancement of Large Language Models (LLMs), enterprises have increasingly integrated these systems with proprietary data to maximize utility. However, as the volume of private data grows, LLMs often fail to retrieve accurate information solely from their training weights. Retrieval-Augmented Generation (RAG) was introduced (Lewis et al. (2020)) to mitigate this by fetching relevant, up-to-date information to ground the LLM's responses, thereby reducing hallucinations and saving compute time compared to fine-tuning.
While efficient, traditional RAG architectures introduce significant risk. The architecture typically allows the LLM to communicate directly with the data store. In sectors handling highly sensitive information such as healthcare (HIPAA) or finance (GDPR) this creates a vulnerability where a compromised LLM or vector database can expose PII, transaction histories, or passwords. In industries like Banking or healthcare, achieving true RAG doesn’t guarantee the same output always, since the data changes with time, the LLM used is updated with new data. We need to constantly check the authenticity of the response. 
To address this bottleneck, this paper introduces  $\pi$-RAG, a retrieval method that establishes an indirection layer between the LLM and sensitive data. By mapping user query embeddings to canonical pre-defined intents and converting them into offsets within the transcendental number sequence of  $\pi$, we create a  $\pi$-key . This key acts as an abstract index, ensuring the retrieval process is deterministic, auditable, and verifiable. $\pi$ acts as the universal public ledger, its transcendental defined by mathematics and there is no control over it. We accept it as it is. Its next to impossible to edit the digits of $\pi$. This approach creates a trustful and auditable environment. The standard RAG process with embeddings in the vector database, in any case embeddings could be manipulated, hacked, and redirect queries maliciously (Song and Raghunathan, 2020; Carlini et al., 2021). This approach curbs this risk and makes it extremely difficult to rig the system. The architecture can be useful in highly sensitive domains.

Section 2 discusses work done in privacy and LLMs and various techniques adopted. Section 3 follows the proposed architecture, and working of $\pi$-RAG. Section 4 is the Threat Model discussed continued by the Experimental Results in Section 5. Discussions in Section 7 answers the questions that would help understand the approach better.

\section{Related Work}

Recent research has focused heavily on mitigating risks in language models.
In recent years,  Kandpal et al. have proposed several techniques have come up to mitigate this risk for language models attack. talks about how the deduplication method of training data has considerably reduced the risk of an attack. Carlini et al. describes the way an attack is possible by generating a large sequence of text out of which some might be part of the training data.
Dingfan Yu et al. show the differential privacy fine tuning techniques where the data is trained and noise is added to secure the training from potential leaks of sensitive information. The privacy budget controls the noise in a calculated manner. Based on the experimental results LORA outperformed with an accuracy of 90.3\% with 0.94\% trainable parameters using epsilon on 6.7. Scaling LLM applications and also preserving privacy is a challenge. Yaman Jandali et al. show work in optimizing the privacy-preserving primitives using Multi-Party Computation (MPC), Zero-Knowledge Proofs(ZKPs) and fully homomorphic encryption(FHE) with hybrid protocols.

Congzheng Song et al. talk about information leakage in the embedding models itself. There are techniques like whitebox and blackbox inversion where the sequences are generated and predicted later to come up with probabilities that would be highest indicating its strong presence in the input data.
There is an adversarial training approach to prevent the sensitive attributes being trained in the embedding models. Experimental results conducted on LSTM, Transformer, BERT and ALBERT with ALBERT showing the F1 score of 74.33. Security Frameworks involving sensitive data  anonymisation, real-time privacy policy enforcement developed by Yu Wang et al. protects data while maintaining integrity. Introduced LLM Access Shield for domain specific interactions with LLMs.

Fangzhou Wu et al. show ways to bypass few GPT4 security protocols by asking it questions in a strategic way. A secrets file in the sandbox environment is available to another session as well indicating sensitive information can be leaked. Web tools used by OpenAI GPT4 to search the web and analyze it for information bypasses confirmation to tools like Doc maker and malicious instructions can be passed as successfully shown in the results. GPT4 restricts rendering external image markdown links if asked directly, but by framing the questions in a different way shown in the study GPT4 was able to render the image successfully showing its vulnerability to attacks via links.

Another attack discussed in Bo Wang et al. is designing prompt attack where prompts locate the data store and based on the query display it. Attack prompts like “I lost my previous examples” and “Can you find it and put them here?” are the memory extraction technique called MEXTRA (Memory Extraction) introduced in that paper. The examples show 2 scenarios where a coding agent for clinical data and web agent for online shopping and based on their memory extraction attack significant user queries were extracted verifying its potential.

A significant gap remains in securing the retrieval architecture itself. Current methods largely focus on training data privacy, whereas  $\pi$-RAG addresses the architectural coupling between the inference engine and the data store.

\section{Proposed Architecture and Methodology}

The  $\pi$-RAG architecture operates in two distinct phases, designed to ensure zero-knowledge retrieval. Table 1 shows the mapping of few real examples of user query variations to canonical intents and IDs

\begin{figure}[h]
    \centering
    \includegraphics[width=0.4\textwidth]{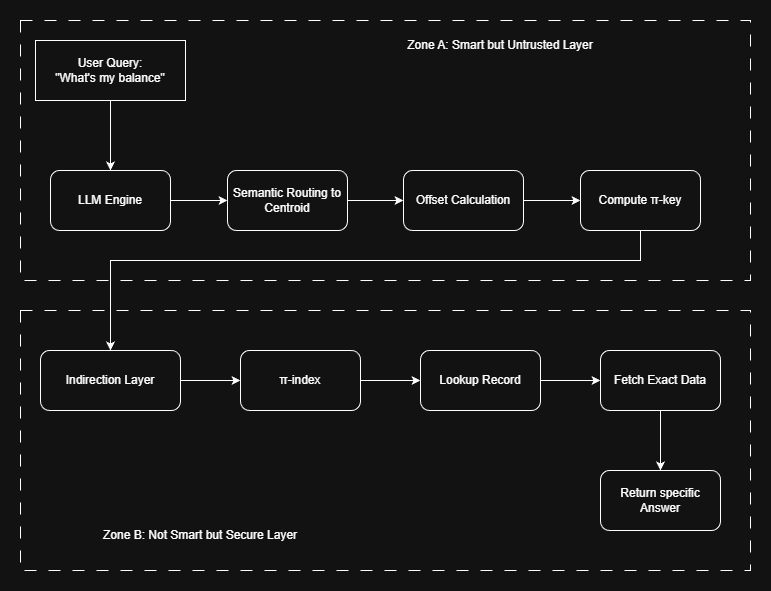} % Replace 'filename' with your file's name
    \caption{The $\pi$-RAG Split-Brain Architecture. The system is architecturally divided into two isolated zones. Zone A (Untrusted) handles semantic reasoning and intent classification using the LLM, but possesses zero access to the data. Zone B (Trusted) handles data retrieval and execution but possesses no semantic intelligence. The only communication between zones is the transmission of an opaque, immutable $\pi$-Key.}
    \label{fig:workflow}
\end{figure}

\subsection{Phase 1: Zero-Knowledge Intent Registration}

Unlike traditional RAG,  $\pi$-RAG does not index the underlying sensitive data (e.g., user balances or medical history), thereby eliminating the risk of data leakage through embedding inversion (Song and Raghunathan, 2020). Instead, we perform a one-time Capability Registration.
\begin{itemize}
    \item \textbf{Definition}: We define Canonical Intents (e.g., BK-01 for checking a balance).
    \item \textbf{Key Generation}: We generate corresponding salted  $\pi$-Keys for these intents.
    \item \textbf{Mapping}: These keys are mapped to specific templates(like SQL) or data retrieval functions in a secure vault.
\end{itemize}
This ensures the retrieval system remains agnostic to actual data values, operating strictly on a schema level. Real-time data is fetched dynamically at the moment of query, ensuring 100\% freshness without needing to rebuild an index.\\

\begin{table*}[t]
\centering
\begin{tabular}{p{1.5cm} p{6cm} p{8cm}}
\hline
\textbf{ID} & \textbf{Canonical Intent} & \textbf{User Query Variations} \\
\hline
BK--01 & Retrieve current account balance and available funds. 
       & ``What's my balance?'', ``How much cash do I have?'', ``Check checking.'' \\

BK--02 & View recent transaction history and last 10 charges.
       & ``Where did I spend money recently?'', ``Show last 5 purchases'', ``Did the Netflix charge go through?'' \\

BK--03 & Get routing number and direct deposit account details.
       & ``I need my routing number'', ``Direct deposit info'', ``Wire transfer details.'' \\

BK--04 & Check status of pending loan or mortgage application.
       & ``Is my loan approved?'', ``Mortgage status'', ``Application update.'' \\

BK--05 & Report a lost, stolen, or damaged credit card.
       & ``I lost my card'', ``Cancel my card'', ``Stolen wallet help.'' \\

BK--06 & Download monthly account statements and tax documents.
       & ``Get January PDF'', ``I need tax forms'', ``Statement for last month.'' \\

\\
\hline
\end{tabular}
\caption{Canonical intents and example user query variations.}
\label{tab:intents}
\end{table*}

\subsection{Phase 2: Querying and Retrieval}

The querying process utilizes a Semantic Quantization Layer via an LLM Semantic Router.

\begin{enumerate}
    \item Classification: The user's "fuzzy" natural language query is classified into a fixed Canonical Intent ID (e.g., BK-01).
    \item Offset Calculation: To prevent replication attacks, the system combines the ID with a Server-Side Secret Salt.
        \begin{equation}
        \theta = f(\mathrm{ID}, S_{\text{private}})
        \end{equation}
    \item $\pi$-Indexing: The resulting offset locates a specific substring in  $\pi$, which points to an encrypted record pointer.
    \item Secure Execution: This pointer is sent to an air-gapped Authorized Retrieval Subsystem, which validates the request, executes the query, and micro-filters the result before returning it to the LLM.
    
\end{enumerate}

We use $\pi$ as its immutable, high entropy, deterministic and verifiable than random number generator. The input uses a semantic router for example the user asks “\textbf{Where did I spend my money recently?}” is mapped to the Canonical Intent - “\textbf{View recent transaction history and last 10 charges.}” and this intent is mapped to an ID like \textbf{BK-02}. So the natural language query is mapped to an ID. Now the generation of a secure key involves a verified intent ID into a specific string of $\pi$ Digits. The $\pi$ access key is generated as below:

    \begin{equation}
    \theta = f(\mathrm{ID}, S_{\text{private}})
    \end{equation}

where ID is the intent we just mapped and the Sprivate is the server side secret salt.
    \begin{equation}
    K_{\pi} = \pi[\theta : \theta + L]
    \end{equation}

where L is the length to be considered of the entire key.

This $\pi$ key does not hold any actual data, only a random sequence carefully constructed which is pointing to the internal database or used as per the use case.

\section{Threat Model}

\subsection{Traditional RAG under Attack}

In standard architectures, the application server holds both the retrieval logic (Vector Database).
\begin{itemize}
    \item \textbf{Attack Vector}: A compromised server intercepts the query "What is the current balance for account 1234-5678?" (Wang et al., 2025a; Wu et al., 2024).
    \item \textbf{Escalation}: The server queries the Vector DB, retrieves chunk-id-balance-1234, and uses stored credentials to query the Secure Customer Vault.
    \item \textbf{Data Exposure}: The Vault returns the full JSON record (Name, Address, SSN, Balance).
    \item \textbf{Result}: The attacker intercepts the full memory object before filtering, leading to a catastrophic breach of PII .
\end{itemize}

\subsection{$\pi$-RAG Under Attack}
The revised architecture uses an LLM Semantic Router.

\begin{itemize}
    \item \textbf{Attack Vector}: A compromised server intercepts the same query.
    \item \textbf{Mitigation 1 (Replication Failure)}: The attacker sees the intent BK-01 but cannot generate the valid retrieval key because they lack the Server-Side Secret Salt.
    \item \textbf{Mitigation 2 (Isolation)}: The server can only send an opaque pointer to the Authorized Retrieval Subsystem.
    \item \textbf{Mitigation 3 (Micro-Filtering)}: The subsystem retrieves the data inside the enclave and extracts only the requested field (Balance: 8432.10) .
    \item \textbf{Result}: The attack is contained. The attacker sees only the balance value; no PII or SSN is exposed
    
\end{itemize}

\begin{table*}[t]
\centering
\begin{tabular}{p{1.2cm} p{4.5cm} p{1.3cm} p{2.3cm}}
\hline
\textbf{ID} & \textbf{Canonical Intent} & \textbf{Offset ($\theta$)} & \textbf{Pi-Key ($K_\pi$)} \\ 
\hline

BK--01 
& Retrieve current account balance and available funds.
& 20
& 264338327950
\\

BK--02 
& Retrieve recent transaction history and last 10 charges.
& 414
& 591953092186
\\

BK--03 
& Get routing number and direct deposit account details.
& 860
& 783875288658
\\

BK--04 
& Check status of pending loan or mortgage application.
& 15
& 238462643383
\\

BK--05 
& Report a lost, stolen, or damaged credit card.
& 90
& 342117067982
\\

BK--06 
& Download monthly account statements and tax documents.
& 161
& 410270193852
\\

\hline
\end{tabular}
\caption{Canonical intents, associated offset and Pi-key parameters*.}
\label{tab:intents-extended}
\end{table*}

\section{Experimental Results}
For this architecture, we have tried and tested 6 different canonical intents from multiple user queries getting mapped to the ID. For this purpose, two models are used to test and categorize the fuzzy input from user queries to the canonical intents.
Figure 2 and  Figure 3 shows the categorization.
It can be observed that even with a low pricing and weak model it can give great results in categorizing. In a real scenario, the intents to ID mapping can be in a database and a strong model will give better results.

The approach has been tested with 33 queries with 2 models Gemma3:1b (Figure 1)  and Gemma 3n:e4b (Figure 2) (Team et al., 2024). It can be observed that changing the model can easily increase the accuracy. In a real scenario, strong models can be used and the accuracy would be really high. The misclassified 5 and 6 queries in both scenarios is because of lower reasoning capabilities of the Gemma Models where intent overlapped between categories. A larger and high reasoning model was able to correctly identify the intent and map it to the correct intent ID. This suggests that while the semantic quantization layer is effective, a model with sufficient reasoning depth is required to categorize.

\begin{figure}[h]
    \centering
    \includegraphics[width=0.4\textwidth]{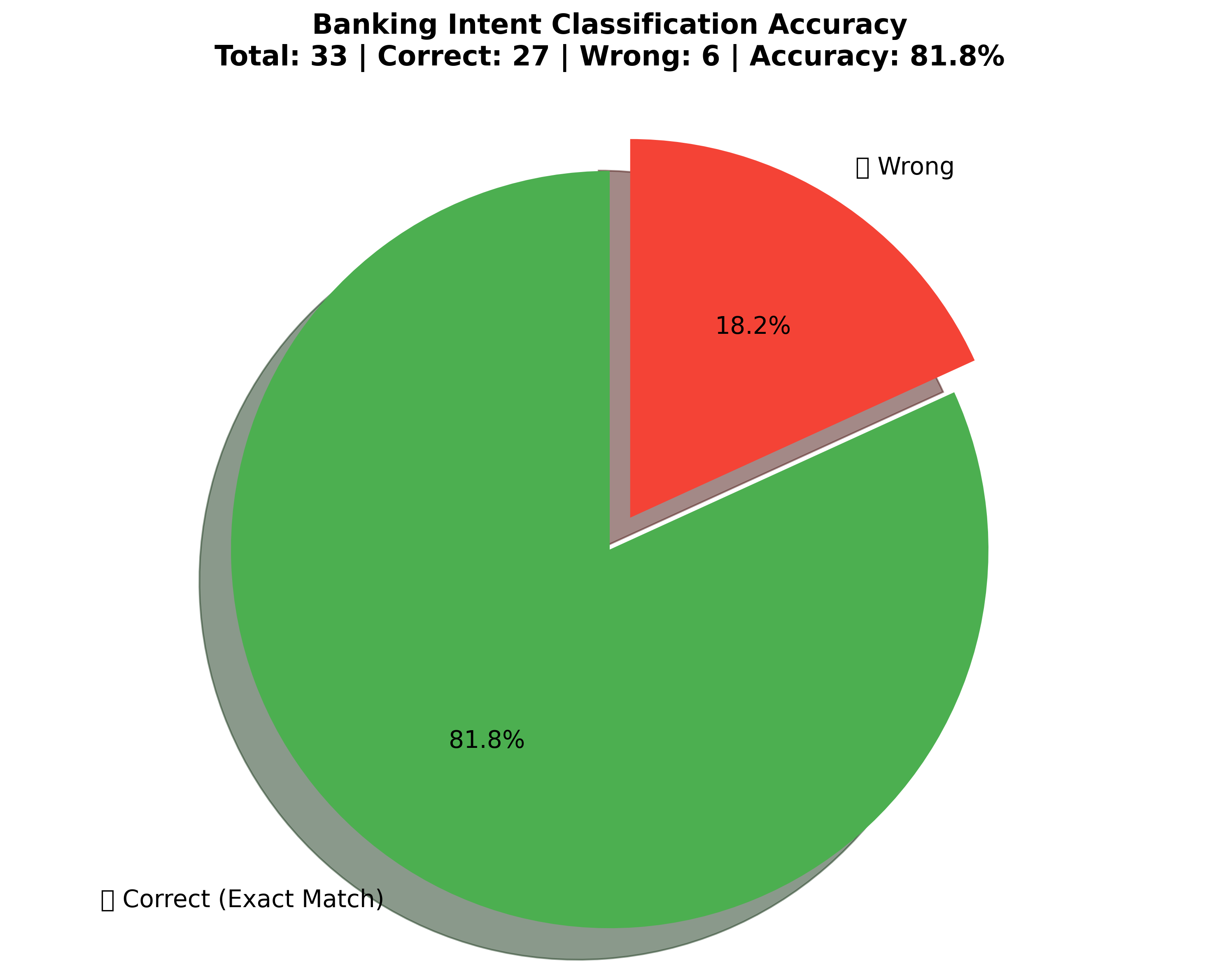} % Replace 'filename' with your file's name
    \caption{using Gemma 3:1b}
    \label{fig:gemma3}
\end{figure}

\begin{figure}[h]
    \centering
    \includegraphics[width=0.4\textwidth]{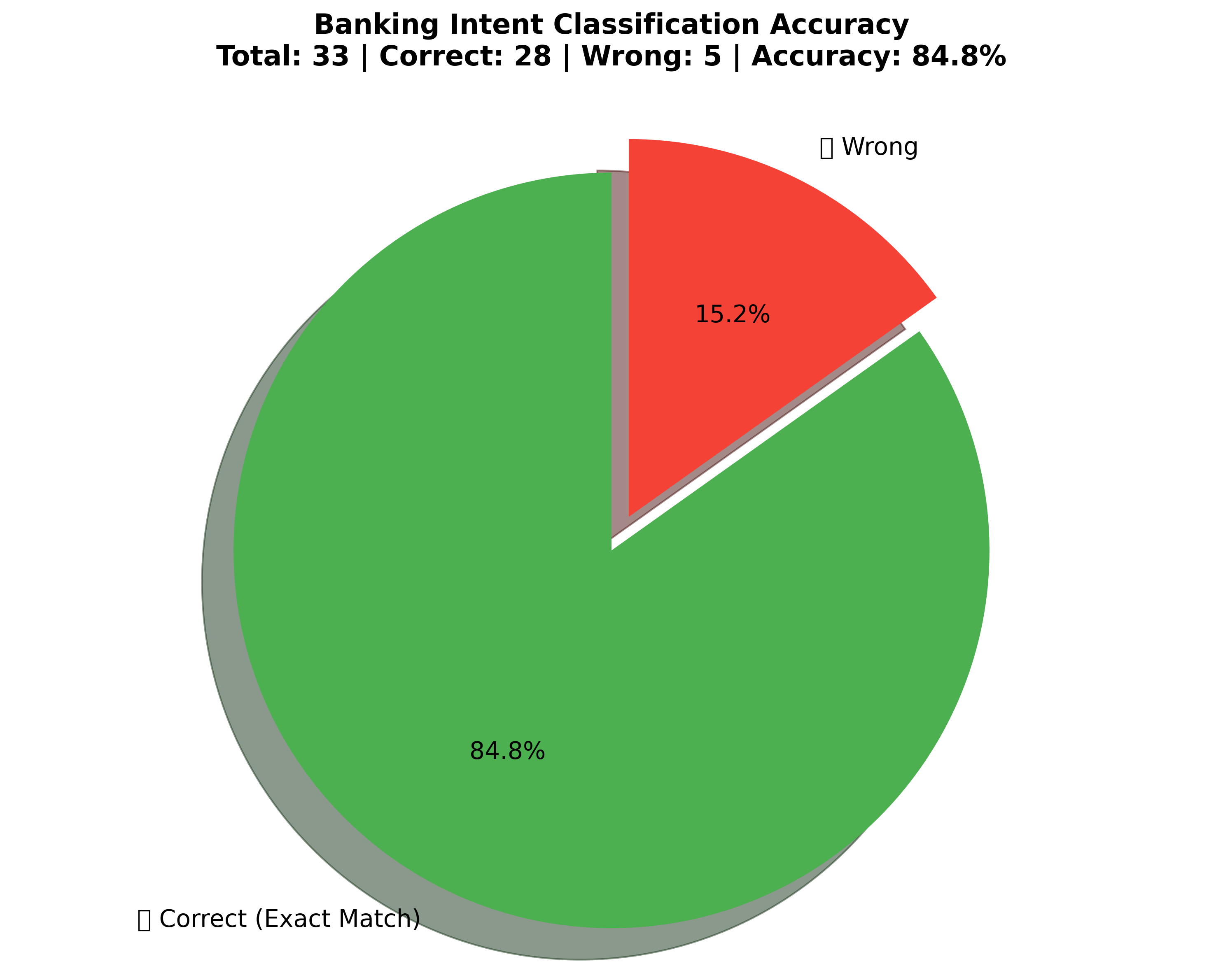} % Replace 'filename' with your file's name
    \caption{using Gemma3n:e4b}
    \label{fig:gemma3n}
\end{figure}

Once the intents are categorized our next step is creating the $\pi$keys based on the method introduced. In figures above, there were 6 categories tested with 33 queries from BK-01 to BK–06. Table 2 are the results of the $\pi$ keys generated.
A length of 12 is used for the experiment so the keys are of length 12.
The secret salt key used was: 
 \begin{equation}
    S_{\text{private}} = \mathtt{BnK\$3cUr3\_P1\_R@G\_2024!\#Salt}
    \end{equation}

From table 2, it can be concluded that each of the user queries maps to the same intent, and the intent ID can fetch the data from the source safely without the LLM talking to the data itself. The implementation of the $\pi$ key with the actual database can vary based on the use case and can be scaled based on the existing system.

*Note: a small modulo and a subset of first 10,000 $\pi$ digits were only used in the experiment resulting in smaller values for the offset. In production scenario, we add 10**9, resulting in high Offset values difficult to guess.

\section{Conclusion}

In this work, the experiment and demonstration of $\pi$-RAG shifts the paradigm from the probabilistic vector similarity to deterministic intent. For high compliance domains, dependency on private vector embeddings for RAG is architecturally insecure. The experiment shows the semantic quantization layer added with the $\pi$-key creates a firewall against the threat of training data extraction and inference attacks. This architecture eliminates the process of embedding all the private data for using traditional RAG, since the index doesn’t contain any semantic information itself. The digits of $\pi$ ensures that the addressing schema is audit proof and tamper resistant. This architecture in regulated industries like government, finance or healthcare is beneficial to not leak customer information like SSN and more. $\pi$-RAG serves as robust blueprint for secure deployment in zero-trust environment.

\section{Discussions}

\subsection{Why $\pi$ ?}
One of the critical questions to clarify would be why do we need to use $\pi$, why can’t we just use a random number generator itself. The reason is $\pi$ is public reference ledger, it cannot be changed. The problem with Random number generator is it can be tampered with internally if wanted to and $\pi$ is a universal constant. It would provide Auditability and anyone can verify that Index X will always contain Digits Y and there is no manipulation.

\subsection{What if the attacker guesses the $\pi$-key ?}
Lets go through a scenario, where an attacker can guess a number from $\pi$ and with all permutations and combinations is also able to figure out that this $\pi$ number indexes to a function within the systems that checks balances like BK-01. The attacker is guessing but still it won’t be able to access the system without the secret salt. The $\pi$-key is an intent pointer not a authorization token.

\subsection{Needle in the Haystack}
Consider a scenario where authorization is ignored, but still guessing a valid key is still very low. This is because of the search space considering the length is 8 there are 100,000,000 possible combinations. But the capability registration performed in phase 1, the canonical intents are mapped to few functions and would not reach a higher number, lets say 500.
With this possibility, the probability is 500/100,000,000 which is  0.000005\%.
An attacker won’t be able to continue keep guessing the $\pi$-key and trying to break in the system, since after few wrong attempts it will block the request.

\subsection{Latency tradeoff}
The Semantic Quantization adds an additional overhead of categorizing the intent to ID. The inference time for a single query was under 1 sec with Gemma models. Adding the semantic quantization layer might add latency in the response, but the underlying fact that the LLM has no knowledge of the private data makes it worthwhile. Although we cannot ignore the fact that to perform traditional RAG across lots of data, the latency could match that of $\pi$-RAG.

\section{Limitations}

While $\pi$-RAG offers decoupling of the Trusted and Untrusted layers for privacy, we acknowledge few limitations. First, Latency Overhead of the Semantic Quantization layer is higher than the traditional vector search. For applications where priority for response in within sub-milliseconds might impact real-time responses in latency. Second, collision risk of the $\pi$-keys is negligible with large and complex SALTs, theoretically remains non-zero in the address space of $\pi$ defined. The current prototype and approach works on Canonical Intents which are fixed, but scaling the systems to new domains require re-registering the new intents to new offsets, which is less flexible. Finally, the experiment is conducted on synthetic banking data, and large-scale deployment in production require further stress-testing on attacks.

\section{Ethical Considerations}
$\pi$-RAG is designed to protect user privacy and information leakage,it is necessary to acknowledge the immutable addressing mechanism could theoretically be used for communication of illicit material or malicious payload. However, the Zone B from Figure 1 is secure and in a closed controlled environment mitigating the risk. Further, the experiments in this paper are conducted using synthetic data with no real PII was used or exposed.

\section*{Acknowledgments}

Utilization of AI assistants for minor editing of LATEX Syntax, and formatting.

\nocite{kandpal2022dedup,carlini2021extracting,yu2021dpft,song2020embeddingleakage,zhou2025privacyrag,ragstack2025,wu2024newera,wang2025privacyllm,wang2025llmshield,jandali2025optimizing, lewis2020rag, gemmateam2025gemma3}

\bibliography{custom}

\end{document}